\documentclass{ws-procs9x6}

\begin{document}

\title{STRUCTURE OF EXOTIC NUCLEI}

\author{
  J. Dobaczewski$^{1-3}$, 
  N. Michel$^{2-4}$,
  \underline{W. Nazarewicz}$^{1-3}$, \\
  M. P{\L}oszajczak$^5$, 
  M.V. Stoitsov$^{2-4,6}$
  }

\address{$^1$Institute of Theoretical Physics, 
Warsaw University,  ul. Ho\.za 69, PL-00681, Warsaw, Poland}  
\address{$^2$Department of Physics and Astronomy, 
The University of Tennessee, 
Knoxville, Tennessee 37996}
\address{$^3$Physics Division, 
Oak Ridge National Laboratory, 
P.O. Box 2008, \\ Oak Ridge, Tennessee 37831}
\address{$^{4}$Joint Institute for Heavy Ion Research,
           Oak Ridge, Tennessee 37831}
\address{$^{5}$ Grand Acc\'{e}l\'{e}rateur National d'Ions Lourds (GANIL), 
CEA/DSM -- CNRS/IN2P3, BP 55027, F-14076 Caen Cedex 05, France}
\address{$^{6}$Institute of Nuclear Research  and Nuclear Energy,
             Bulgarian Academy of Sciences, Sofia-1784, Bulgaria}

%%%%%%%%%%%%%%%%%%%%%%%%%%%%%%%%%%%%%%%%%%%%%%%%%%%%%%%%%%%%%%
% You may repeat \author \address as often as necessary      %
%%%%%%%%%%%%%%%%%%%%%%%%%%%%%%%%%%%%%%%%%%%%%%%%%%%%%%%%%%%%%%

\maketitle

\abstracts{The progress in the modeling of exotic nuclei 
with an extreme neutron-to-proton ratio is discussed. 
Two topics are emphasized: (i) the quest for
 the universal microscopic nuclear 
energy density functional and (ii)  the progress in
the continuum shell model.
}

\section{Introduction}
The goal of nuclear structure theory is to build a unified
 microscopic framework in which bulk nuclear properties
(including masses, radii, and moments, structure of nuclear matter),
nuclear excitations (including a variety of collective phenomena), and
nuclear reactions can all be described. While this goal is
extremely ambitious, it is no longer a dream. 
Indeed,
hand in hand with experimental developments in the
radioactive nuclear beam (RNB) experimentation, a qualitative change in
theoretical modeling is taking place. Due to the influx of new ideas and
the progress in computer technologies and numerical algorithms, nuclear
theorists have been quite successful in solving various pieces of the
nuclear puzzle. 

During recent  years, we have witnessed  substantial progress in many
areas of theoretical nuclear structure. The Effective Field Theory
(EFT)  has
enabled us to construct high-quality NN and NNN bare interactions
consistent with the  chiral symmetry of QCD\cite{[Ent03],[Glo03a]}. New
 effective interactions in the medium have been developed which, together
with a powerful suite of {\em ab-initio} approaches, provide a
quantitative description of light
 nuclei\cite{[Pie02],[Wir02],[Bar02],[Nav03],[Kow03]}.
  For heavy systems, {\em
global} modern shell-model 
approaches\cite{[Cau02],[Hon02],[Lan02],[Lan03]} 
and self-consistent
mean-field methods\cite{[Gor02],[Ben03],[Sto03]} offer a level of
accuracy typical of phenomenological approaches based on parameters {\em
locally} fitted to the data. By exploring connections between models in
various regions of the chart of the nuclides, nuclear theory aims to
develop  a comprehensive theory of the nucleus across the
entire nuclear landscape. 

From a theoretical point of view, short-lived exotic nuclei far from
stability offer a unique test of those aspects of the many-body theory 
that depend on the isospin degrees of freedom\cite{[Dob97a]}. The
challenge to microscopic theory is to develop methodologies to reliably
calculate and understand the origins of unknown properties of new
physical systems, physical systems with the same ingredients as familiar
ones but with totally new and different properties.  The hope is that
after probing  the limits of extreme isospin, we can later go back  to
the valley of stability and   improve the description of normal nuclei.

\section{Towards the Universal Nuclear Energy Density Functional}

For medium-mass and heavy nuclei, a critical challenge is the quest for the universal
energy density functional, which will be able to describe properties of
finite nuclei (static properties, collective states, large-amplitude
collective motion)  as well as  extended asymmetric nucleonic matter (e.g.,
as found in neutron stars).
Self-consistent  methods based on the density functional theory 
(DFT) have  already
achieved a level of sophistication and precision which allows analyses
of experimental data for a wide range of properties and for arbitrarily
heavy nuclei. For instance,   self-consistent Hartree-Fock (HF) and  
Hartree-Fock-Bogoliubov (HFB)  models
are now able to reproduce  measured  nuclear  binding energies  with an
impressive rms error of $\sim$700 keV \cite{[Gor02],[Sam02],[Gor03]}. However,
much work  remains to be done. Developing a universal nuclear density
functional  will require a better
understanding of the    density dependence, isospin effects, 
and pairing,
as well as an improved treatment of symmetry breaking effects and
  many-body correlations. 
  
\subsection{Density Functional Theory and Skyrme HFB}  
 
The density functional theory\cite{[Hoh64],[Koh65a]} has been
 an extremely successful approach for the
description of ground-state properties of bulk  
(metals, semiconductors, and insulators) and complex 
(molecules, proteins, nanostructures)
materials. It has also been used with great success in nuclear 
physics\cite{[Mig67],[Neg70],[Neg72],[Bra97]}.
The main idea of DFT is to describe an interacting system of fermions
via its densities and not via its many-body wave function. 
The  energy of the many body system can be written as 
a density functional, and the ground state energy is obtained through the
variational procedure. 

The nuclear energy density functional appears naturally in the Skyrme-HFB
theory\cite{[Sky59],[Vau72]},
or in the local density approximation 
(LDA)\cite{[Neg72],[RS80]}, in which the  
functional depends only on local
densities, and on local densities built from derivatives up to the
second order. In practice, a number of local densities are introduced:
nucleonic  densities, kinetic densities, spin densities,
spin-kinetic densities, current  densities,
tensor-kinetic densities, and spin-current densities. If 
pairing correlations are considered, the number of local densities 
doubles
since one has to consider both particle and pairing densities.

In the case of the Skyrme  effective interaction, as well as
in the framework of the LDA, the energy functional
 is a three-dimensional spatial integral
of local energy density that is a real, scalar,
time-even, and isoscalar function of local densities and their first
and second derivatives.  In the case of no proton-neutron
mixing, the construction of the most general energy density
that is quadratic in one-body local densities can be found
 in Ref.\cite{[Dob96b]}. With the proton-neutron
mixing included, the construction can be performed in an analogous
 manner\cite{[Per03]}.

\subsection{From finite nuclei to bulk nucleonic matter}

In the limit of the infinite  nuclear matter, the density functional 
is reduced to the 
nuclear equation of state (EOS). The EOS plays a central role in nuclear
structure and in heavy-ion collisions. It also determines the static
and dynamical behavior of stars, especially in supernova explosions and
in neutron star stability and evolution. Unfortunately, our knowledge of
the EOS, especially at high densities and/or temperatures, is very poor.
Many insights about the density dependence of the EOS, in particular the density dependence of the
symmetry energy, 
can be obtained from microscopic calculations of neutron matter using
realistic nucleon-nucleon forces\cite{[Fri81],[Mor02],[Car02]}. 
Those results will certainly be helpful when constraining 
realistic energy density
functionals. Another constraint comes form 
measurements of neutron skin and radii\cite{[Hor01],[Fur02]}. 
Recently,  a correlation between the neutron skin in heavy nuclei
and the derivative of the neutron equation of state has been 
 found\cite{[Typ01],[Fur02],[Die03]},  which
provides a way of giving a stringent constraint on the EOS if the 
neutron radius of a heavy nucleus is measured with sufficient accuracy.

A serious difficulty when extrapolating from finite nuclei
to the extended nuclear matter is due to the diffused neutron surface
in neutron-rich nuclei. As discussed in Ref.\cite{[Dob02c]},
the nuclear surface cannot simply be
regarded as a {\it layer of nuclear matter at low density}. In this zone
the gradient terms  are as important in
defining the energy relations as those depending on the local
density.

\subsection{The First Step: Microscopic Mass Table}

\begin{figure}[htb]
  \begin{center}
 \leavevmode
  \epsfxsize=7.5cm
  \epsfbox{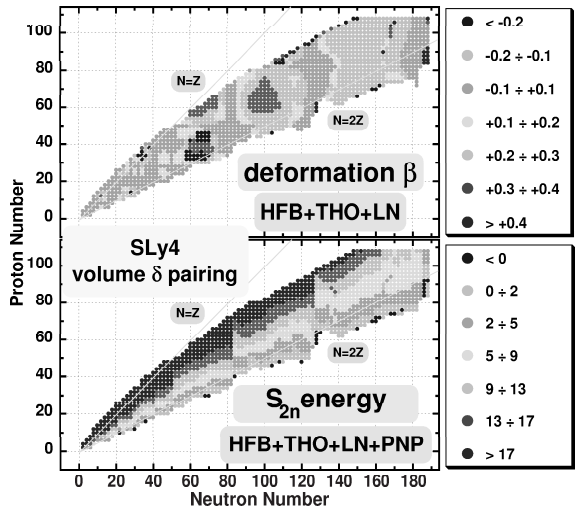}
\caption{Quadrupole deformations $\beta$ (upper panel) and
two-neutron separation energies $S_{2n}$ in MeV (lower panel) of
particle-bound even-even nuclei calculated within the HFB+THO method
with Lipkin-Nogami correction followed by exact particle number
projection. The Skyrme SLy4 interaction and volume contact pairing
were used. (From Ref.~\protect\cite{[Sto03]}.)}
\label{deforHFB}
\end{center}
\end{figure}

Microscopic mass calculations  require a simultaneous description of
particle-hole, pairing, and continuum effects -- the challenge that only
very recently could be addressed by mean-field methods. A new 
development\cite{[Sto03]}
is the solution of deformed HFB equations 
by using the local-scaling point
transformation \cite{[Sto98b],[Sto99]}. A representative example of 
deformed HFB calculations,  recently implemented using the parallel
computational facilities at ORNL, is given in Fig.~\ref{deforHFB}. 
By creating a simple load-balancing routine that allows one to scale the
problem to 200 processors, it was possible to calculate the entire deformed
even-even mass table in a single 24 wall-clock hour run (or approximately
4,800 processor hours).

Future calculations will take into account a number of 
improvements, including
(i) implementation of the exact particle
number projection before variation\cite{[She00a]}; (ii)
better modeling  of the density dependence of the effective interaction
by considering 
corrections beyond the mean-field and
three-body effects\cite{[Dug03]}, the
surface-peaked effective mass\cite{[Far01],[Gor03]},
 and 
better treatment of pairing\cite{[Dob02c]};
(iii)
proper treatment of the time-odd fields \cite{[Ben02]};
and (iv)
inclusion of dynamical zero-point fluctuations  
associated with the nuclear collective
motion \cite{[Rei78],[Rei87],[Rei99]}.
As far as the density dependence is concerned,
many insights can be obtained from the EFT\cite{[Pug03]}.
The resulting universal
energy density functional will be fitted to nuclear masses, radii, 
giant vibrations,
and other global nuclear characteristics.

Finally, let us remark that a realistic energy density
functional does not have to be related to any given effective force.
This creates a problem if a symmetry is spontaneously broken.
While the projection can be carried out in a straightforward manner
for  energy functionals that  are related to a two-body potential,
the restoration of  spontaneously broken
symmetries of a general density functional  poses a  conceptional
dilemma\cite{[Goe93],[Per95]}.

\section{Continuum Shell-Model}

The major theoretical challenge in the microscopic
description of  nuclei, especially weakly bound ones, 
is the rigorous treatment of both the many-body correlations 
and the continuum of positive-energy states and decay channels. 
The importance of continuum for the description of resonances is obvious.
 Weakly bound states cannot be
described within the
closed quantum system formalism since there always appears a virtual
scattering into the continuum phase space  involving intermediate
scattering states. 
The consistent treatment of continuum in multi-configuration mixing
calculations is the domain of the continuum shell model (CSM)
(see Ref.\cite{[Oko03]} for a review).
 In the following, we briefly mention one 
recent development in the area of the CSM, the so-called Gamow Shell Model.

\subsection{Gamow Shell Model}

Recently, the multiconfigurational CSM
in the complete Berggren basis, the so-called Gamow Shell Model (GSM), has
been formulated  \cite{[Mic02],[Mic03]}. 
The s.p. basis of GSM is given
by the Berggren ensemble \cite{[Ber68]} which contains 
Gamow states (or resonant
states and the non-resonant continuum).
The resonant states are the generalized eigenstates of the
time-independent Schr\"odinger equation which are regular at the origin and
satisfy purely outgoing boundary conditions. 
They correspond to the poles of the $S$ matrix in the complex
energy plane lying on or below the positive real axis.
 
There exist several completeness relations involving resonant states
\cite{[Lin93]}.
In the heart  of GSM is
the Berggren completeness relation:
\begin{equation}
\label{eq1x}
\sum_n|u_n\rangle\langle{\tilde{u}_n}| + 
\int_{L_+}|u_k\rangle\langle{\tilde{u}_k}|dk = 1 ~ \ ,
\end{equation}
where $|u_n\rangle$ are the Gamow states (both bound states and the decaying resonant
states lying between the real $k$-axis and the complex contour $L_+$) and 
$|u_k\rangle$ are the scattering states on $L_+$.
(For neutrons, 
$l=0$ resonances do not exist and, sometimes, one has to include the
anti-bound  $l=0$ state 
in the Berggren completeness relation \cite{[Bet03],[Hag03]}. This implies a
modification of the complex contour $L_+$, which has to enclose
this pole.)
 As a consequence of the analytical continuation, the resonant states
 are normalized according to the
squared radial wave function and not to the modulus of the squared radial
wave function. In practical applications, one has to discretize
the integral in (\ref{eq1x}). Such a discretized Berggren relation is formally
analogous to the standard completeness relation in a discrete basis of 
$L^2$-functions and, in the same way, leads to the eigenvalue problem
$H|{\Psi}\rangle=E|{\Psi}\rangle$. However, as the formalism of Gamow states is non-hermitian,
the matrix $H$ is complex symmetric. 

One of the main challenges in the CSM is the determination of   many-body
resonances because of a huge number (continuum) of surrounding 
many-body scattering states. 
A practical solution to this problem 
 has been proposed in
Refs.\cite{[Mic02],[Mic03]}. It is based on the fact that resonances have 
significant overlap with many-body states calculated in the pole approximation in
which  the Hamiltonian is diagonalized in a smaller basis
consisting of s.p.  resonant states only. 
 The eigenstates representing the non-resonant background tend 
 to align along regular trajectories in the complex energy plane.
 As discussed in Refs. \cite{[Bet02],[Bet03]}, the shapes of these 
 trajectories directly
 reflect the geometry of the contour in the complex $k$-plane. In the 
 two-particle case, this information can be directly 
 used to identify the resonance states. However, this is no longer the
  case if more than
 two particles are involved.
 
In the shell-model calculations with Gamow states, only radial 
matrix elements are treated differently 
as compared to the standard shell model.
This means  that  the angular momentum and isospin algebra 
 do not change in the GSM. However, expectation values of operators in 
 the many-body GSM states have both real and imaginary parts.
As discussed in Refs.\cite{[Ber96a],[Bol96],[Civ99]}, 
the imaginary part gives the
 uncertainty  of the average  value. It is also worth noting that, 
 in most cases,
the real part of the matrix element is 
 influenced by the interference with the non-resonant background.

Contrary to the traditional shell model, 
the effective interaction of GSM cannot be represented as 
a single matrix calculated for all nuclei in a given region. 
The GSM Hamiltonian contains  a real 
effective two-body force expressed in terms of space, spin, and 
isospin coordinates.
The matrix elements
involving continuum states are strongly system-dependent and they 
fully  take into account the spatial extension of s.p. wave functions.

\begin{figure}[htb]
  \begin{center}
 \leavevmode
  \epsfxsize=7.5cm
  \epsfbox{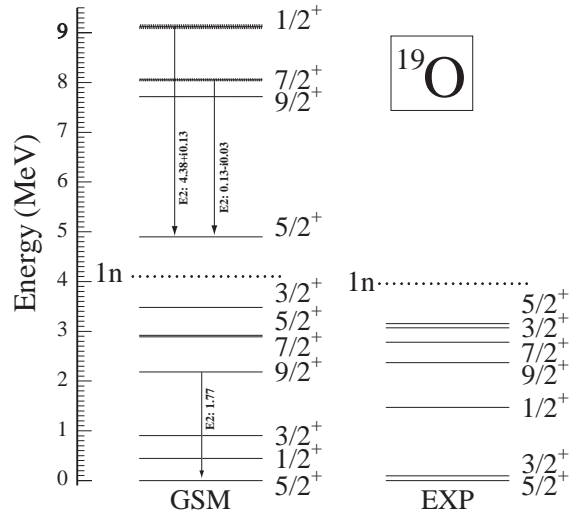}
\caption{
The GSM  level scheme of $^{19}$O 
calculated in the full $sd$ space of Gamow states and employing
the discretized (10 points) $d_{3/2}$ non-resonant continuum. 
The dashed lines indicate 
experimental and calculated one-neutron emission thresholds.
As the number of states becomes large above 
  the one-neutron emission threshold,
  only selected resonances are shown.
Selected E2 transitions
are indicated by arrows and the calculated E2 rates
   (all in W.u.) are given (from Ref.~\protect\cite{[Mic03]}).
}
\label{spectre_O19}
\end{center}
\end{figure}

In the first applications of the GSM, a schematic zero-range surface
delta force was taken 
as a residual interaction.
As a typical example,
the calculated level scheme  of $^{19}$O  is displayed in 
Fig.~\ref{spectre_O19} together with
the selected
E2 transition rates. 
 It is seen that the electromagnetic transition rates
involving unbound states are complex.

The first applications of the GSM to the oxygen and helium isotopes 
look very promising\cite{[Mic02],[Mic03]}. 
The beginning stages of a broad research program has begun which
involves applications of GSM to halo nuclei, particle-unstable nuclear
states, reactions of astrophysical interest, and a variety of nuclear
structure phenomena. The important step will be to develop effective
finite-range interactions to be used in the GSM calculations. One would
also like to optimize the path of integration representing the
non-resonant continuum. In order to optimize
the GSM configuration space, we intend to carry out GSM calculations
 in the Hartree-Fock  basis. To this end, 
a Hartree-Fock program in the Gamow
basis has been developed (GHF) \cite{[Mic03a]}. The GHF method will also
 be applied to describe nuclear vibrational states in the 
 continuum RPA (or QRPA) framework.

\section{Conclusions}

The main objective of this presentation
was to discuss the opportunities in nuclear structure 
that have been enabled by studies of exotic nuclei with 
extreme neutron-to-proton ratios.
New-generation data  will be crucial in pinning down a number
of long-standing questions  related to the effective Hamiltonian,
nuclear collectivity, and properties of nuclear excitations.

One of the major challenges is to develop the ``universal"
nuclear energy density functional that 
will describe properties of finite nuclei as well as extended
 asymmetric nucleonic matter as found in neutron stars.
 This quest is strongly  driven by new data 
 on nuclei far from stability, where new features,
 such as weak binding and altered interactions, make extrapolations of
 existing models very unreliable.
 
Another major task is to 
tie nuclear structure directly to nuclear reactions within a coherent
framework applicable throughout the nuclear landscape.
From the nuclear structure perspective, the continuum shell model is 
the tool of choice that will be able to describe  new 
phenomena in discrete/continuum spectroscopy of exotic nuclei.

\section*{Acknowledgments}
This work was supported in part by the U.S.\ Department of Energy
under Contract Nos.\ DE-FG02-96ER40963 (University of Tennessee) and
 DE-AC05-00OR22725
with UT-Battelle, LLC (Oak Ridge National Laboratory), by
the Polish Committee for Scientific Research (KBN),
%under Contract No.~5~P03B~014~21, 
and by the National Science Foundation Contract No.
0124053 (U.S.-Japan Cooperative Science Award).

%\bibliography{jacwit18,niigata}
%\bibliographystyle{unsrt}

\end{document}